\documentclass[12pt]{iopart}
\usepackage{iopams} 
\usepackage[utf8]{
inputenc}

\usepackage[titletoc,toc,title,page]{appendix}

\usepackage[italian,english]{babel}

\usepackage{microtype} 

\usepackage[normalem]{ulem} 

\usepackage{parskip}
\usepackage[colorlinks=true]{hyperref}
\hypersetup{
  colorlinks   = true, 
  urlcolor     = green!80!black, 
  linkcolor    = blue, 
  citecolor    = red!80!black 
}

\usepackage{graphicx}

\usepackage[usenames,dvipsnames,table]{xcolor}

\graphicspath{{./figures/}}

\newcommand{\norm}[1]{||#1||}

\definecolor{LightCyan}{RGB}{234,250,255}
\definecolor{LightRed}{RGB}{250,234,234}

\usepackage{cite}

\begin{document}

\title{Quantum Extreme Learning of molecular potential energy surfaces and force fields}

\author{Gabriele Lo Monaco}
\let\comma,
\address{Universit\`a degli Studi di Palermo\comma{} Dipartimento di Fisica e Chimica - Emilio Segr\`e\comma{} via Archirafi 36\comma{} I-90123 Palermo\comma{} Italy}
\author{Marco Bertini}
\let\comma,
\address{Universit\`a degli Studi di Palermo\comma{} Dipartimento di Fisica e Chimica - Emilio Segr\`e\comma{} via Archirafi 36\comma{} I-90123 Palermo\comma{} Italy}
\author{Salvatore Lorenzo}
\let\comma,
\address{Universit\`a degli Studi di Palermo\comma{} Dipartimento di Fisica e Chimica - Emilio Segr\`e\comma{} via Archirafi 36\comma{} I-90123 Palermo\comma{} Italy}
\author{G. Massimo Palma}
\address{Universit\`a degli Studi di Palermo\comma{} Dipartimento di Fisica e Chimica - Emilio Segr\`e\comma{} via Archirafi 36\comma{} I-90123 Palermo\comma{} Italy}
\address{NEST\comma{} Istituto Nanoscienze-CNR\comma{} Piazza S. Silvestro 12\comma{} 56127 Pisa\comma{} Italy}


\begin{abstract}
\noindent Quantum machine learning algorithms are expected to play a pivotal role in quantum chemistry simulations in the immediate future. One such key application is the training of a quantum neural network to learn the potential energy surface and force field of molecular systems. We address this task by using the quantum extreme learning machine paradigm. This particular supervised learning routine allows for resource-efficient training, consisting of a simple linear regression performed on a classical computer. We have tested a setup that can be used to study molecules of any dimension and is optimized for immediate use on NISQ devices with a limited number of native gates. We have applied this setup to three case studies: lithium hydride, water, and formamide, carrying out both noiseless simulations and actual implementation on IBM quantum hardware. Compared to other supervised learning routines, the proposed setup requires minimal quantum resources, making it feasible for direct implementation on quantum platforms, while still achieving a high level of predictive accuracy compared to simulations. Our encouraging results pave the way towards the future application to more complex molecules, being the proposed setup scalable.
\end{abstract}

\maketitle

\section{Introduction}
The simulation of complex phenomena involving large biochemical systems, such as protein folding or enzyme activation, requires accurate predictions of energies and inter-atomic forces for different molecular conformations. Standard techniques such as molecular dynamics (MD) require \emph{on-the-fly} computing the energy and inter-atomic forces of the target biochemical system at each step of the simulation. In most MD-like schemes, the Potential Energy Surface (PES) and Force Fields (FF) are determined on empirical grounds or on databases. In contrast, a higher degree of accuracy demands an {\it ab initio} calculation of the PES, typically using density functional theory (DFT) \cite{car1985unified}. The use of DFT on the fly leads to a much longer simulation running time and, due to its huge computational cost, it turns out to be a viable route only for short-time simulations (picoseconds) of small molecules.

A precise knowledge of the functional relation between the molecular geometry and the corresponding energy on the PES, predicted by the DFT, would incredibly speed up ab-initio MD simulations. As Behler and Parrinello observed in \cite{behler2007generalized}, this functional relation may be well approximated by neural networks (NNs) after appropriate training on the pairs of molecular conformations and their DFT energies (or forces). The generation of the PES and FFs using various machine learning (ML) paradigms has been further investigated in \cite{behler2011atom,rupp2012fast,behler2016perspective,gastegger2018wacsf,imbalzano2018automatic,cheng2020evidence,unke2021machine,krems2019bayesian,asnaashari2021gradient,jasinski2020machine,cui2016efficient,dai2023neural,blank1995neural,guan2021structure}. 
Given the intrinsic quantum nature of the PES, it is natural to believe that quantum algorithms may indeed help \cite{ollitrault2021molecular,sajjan2022quantum}. Through the implementation of a supervised learning setup where the role of the classical NN is played by a parameterized quantum circuit (PQC), the authors of \cite{kiss2022quantum,dai2022quantum} provided the first evidence that quantum machine learning (QML) routines may offer improved and more accurate PES predictions. 

Since PQCs can be implemented on NISQ devices, they have been extensively utilized in recent times for different purposes. However, they are not the unique viable QML paradigm for supervised learning. An attractive alternative is provided by quantum extreme learning machines (QELMs). This setup may be advantageous for various reasons. First, it requires reduced training efforts, as in the QELM paradigm no parameters in the quantum platform need to be tuned and the training consists only of a simple classical linear regression. Second, quantum extreme learning machines can be implemented on different physical platforms \cite{fujii2017harnessing,chen2019learning,nakajima2019boosting,martinez2020information,chen2020temporal,kutvonen2020optimizing,martinez2021dynamical,nokkala2021gaussian} and not necessarily on gate-based quantum devices. We propose to use QELM to speed up ab initio molecular dynamics computations. We show that a QELM trained on small sets of geometry-energy pairs is able to efficiently predict the energy of a set of test geometries with great accuracy.\\
The QELM setup that we propose significantly reduces the quantum resources required for the training process compared to other supervised learning paradigms such as VQE (as discussed in \cite{kiss2022quantum}). By quantum resources, we refer to either the depth of the quantum circuits employed, as reported in \Tref{tab:results} and \Tref{tab:vqe} and the number of runs on the quantum platform. In fact, in QELM the number of runs only depends on the size of the training set and the chosen shot statistics; the optimization stage only consists of simple linear regression on the output of the quantum circuit, and no parameter on the quantum device needs to be optimized. In VQE, instead, the number of runs also depends on the number of iterations necessary for the optimization to converge and on the number of parameters in the quantum circuit to be optimized.

\Sref{sec:QEML} contains a self-consistent review of the QELM paradigm. The details about our particular realization of QELM adapted to PES and FF generations are all collected in \sref{subsec:datasets} and \sref{subsec:QELM_setup}. Finally, in \sref{subsec:res}, we collect all the results about the simulations and quantum-hardware realization of the QELM training in the case of lithium hydride, water, and formamide.

\section{Basics of QELM} \label{sec:QEML}

\subsection{QELM as supervised training of a POVM outcome}
The QELM paradigm with classical input data is a supervised learning model. Its task is to \emph{learn} a function $f: \mathbb{R}^{\mathcal{X}}\rightarrow \mathbb{R}^{\mathcal{Y}}$ that well approximates the actual functional relation between inputs and outputs of some training set $\{\boldsymbol{x}^{\rm tr}_{i},\boldsymbol{y}^{\rm tr}_{i}\}_{i=1}^{M_{\rm tr}}$. The quality of the training is assessed on a separate test set of previously unseen pairs $\{\boldsymbol{x}^{\rm test}_i,\boldsymbol{y}^{\rm test}_i\}_{i=1}^{M_{\rm test}}$.

In the case at hand, the training set consists of the coordinates parameterizing the conformations of a given molecular system and the corresponding energy and forces, computed classically with standard quantum chemistry routines via Hartree-Fock (HF) approximation or DFT. While a possible accuracy threshold may be the chemical accuracy $1.6\cdot 10^{-3}\,{\rm Hartree}$, it should be emphasized that the chemical accuracy plays a role only in the comparison with experimental results, where energy differences related to chemical processes are measured, and not in the evaluation of the absolute energy of a molecular geometry obtained through calculations. Even if chemical accuracy is a good target precision to keep in mind, the goodness of a learning training should be actually assessed on the typical fluctuations of the variable $\boldsymbol y$ in the test dataset. If the error is well below such typical fluctuations, we have a good chance to accurately predict the real difference between the energy or the force of two given sampled geometries.  

In a QELM, the classical $\boldsymbol x$ data are encoded in the quantum state of some \emph{input qubits}, initially prepared in a state $\rho_0$. The encoding is performed by applying a parameterized unitary transformation:
\begin{equation}
    \rho_{\boldsymbol{x}}\,=\,U(\boldsymbol x)\rho_{0}\,U^\dagger(\boldsymbol x)
\end{equation}
After such encoding, a POVM measurement $\boldsymbol \mu=\{\mu_a\}_{a=1}^{\Sigma}$ is performed on the input qubits.

Each outcome $a$ is measured with probability
\begin{equation}
\label{eq:probabilities}
    p_a(\boldsymbol x)=\Tr(\mu_a\,\rho_{\boldsymbol x})\,.
\end{equation}
In the QELM protocol, the function $f$  is approximated as:
\begin{equation}
\label{eq:f}
    f(\boldsymbol{x})\,=\,W\boldsymbol{p}(\boldsymbol x)\,,
\end{equation}
where $W$ is a $\mathcal{Y}\times \Sigma$ matrix. The optimization of the unknown matrix $W$ in order to fit the training data can be performed using standard techniques and it consists of a simple linear regression.
\begin{equation}
    W=\boldsymbol{Y}_{\rm tr}\,\boldsymbol{P}_{\rm tr}^+
\end{equation}
where $\boldsymbol P_{\rm tr}$ is a $\Sigma\otimes M_{\rm tr}$ matrix whose columns are the probability vectors $\boldsymbol{p}(\boldsymbol x_i^{\rm tr})$ of the training set. The superscript $+$ denotes the Moore-Penrose pseudoinverse and $\boldsymbol Y_{\tr}$ is a $\mathcal{Y} \otimes M_{\rm tr}$ matrix whose columns are the output $\boldsymbol y_{\rm tr}$ of the training set. The matrix $W$ is thus the one minimizing the root mean square error (RMSE) on the training set $\epsilon_{\rm tr}=\norm{\boldsymbol{Y}_{\rm tr}-W\boldsymbol{P}_{\rm tr}}_{F}/\sqrt{M_{\rm tr}}$ where $\norm{\cdot}_{F}$ is the Frobenius norm. What is relevant is the root mean square error on the test set:
\begin{equation}
\epsilon_{\rm QELM}=\frac{1}{\sqrt{M_{\rm test}}}\norm{\boldsymbol{Y}_{\rm test}-W\boldsymbol{P}_{\rm test}}_{F}\,.
\end{equation}
This is an estimate of the ability of the trained QELM to provide accurate predictions on a new set of test input data. As already mentioned, the root mean square error alone is not a good quantifier and it should be compared with the typical fluctuations in the output values $\boldsymbol y$ if we sample two random elements from the dataset. This scale is measured by the variance
\begin{equation}
{\rm var}(\boldsymbol Y)=\frac{1}{M_{\rm test}}\norm{\boldsymbol Y_{\rm test}-\overline{\boldsymbol Y}_{\rm test}}^2_{F}\,,
\end{equation}
where $\overline{\boldsymbol Y}_{\rm test}$ is the $\cal{Y}\times M_{\rm test}$ matrix whose column are all equal to the mean value $\overline{\boldsymbol y}_{\rm test}$.  If $\epsilon_{\rm QELM}$ is much smaller of this rooted variance, this means that we will able to predict the variation of the variable $\boldsymbol y$ with great precision.

Let us point out that it is always possible to rewrite \eref{eq:f} in the following way:
\begin{equation}
\label{eq:tomography}
    f(\boldsymbol x)=W\boldsymbol{p}(\boldsymbol x)=\Tr\left(\tilde{\boldsymbol \mu}\rho_x\right)
\end{equation}
where $\tilde{\boldsymbol \mu}=W\boldsymbol \mu$ is a vector of $\mathcal{Y}$ operators acting on the Hilbert space of the input qubits. This means that we are actually approximating the function $f$ as a vector of expectation values. 
On real quantum platforms, the probabilities \eref{eq:probabilities} are estimated performing a finite number of shots (\emph{finite statistics}). This implies that a statistical error is always present and how to minimize it is an open problem of crucial relevance \cite{innocenti2023potential, innocenti2023shadow, domingo2022optimal,thanasilp2022exponential}. 
Furthermore, the choice of the embedding unitaries $U(\boldsymbol x)$ and the choice of POVM $\boldsymbol \mu$ strongly affects the performances of the QELM both at infinite and finite statistics.

\subsection{Implementing informationally complete POVMs with a reservoir}
As we have pointed out, QELM is equivalent to estimating the expectation value of some observable. Looking at \eref{eq:tomography}, this is possible only if such observable can be decomposed as a linear combination of the effect of the POVM and this is possible for any operator only if the the POVM is informationally complete (IC). From an experimental point of view, implementing an IC-POVM is in principle a hard task. To overcome this issue, one may let the input qubits interact with some ancillae (the \emph{reservoir}) prepared in a fixed state $\rho_R$. Then, an easy-to-implement measurement  $\boldsymbol \nu$ is performed. For instance, some qubits can be measured in the computational basis. The string of probabilities obtained with this procedure can be written as:
\begin{equation}
    p_a(\boldsymbol{x})=\Tr\left(\nu_a\,\mathcal{U}\rho_{\boldsymbol x}\otimes \rho_R\,\mathcal{U}^\dagger\right)\,=\,\Tr\left(\Phi^\dagger(\nu_a)\rho_x\right)\,,
\end{equation}
where we defined the quantum channel $\Phi(\rho)=\mathcal{U}\rho\otimes \rho_R\mathcal{U}^\dagger$ and $\Phi^\dagger$ denotes its dual. Comparing this expression with \eref{eq:probabilities}, it is evident that the role of the interaction of the input qubits with the reservoir is to generate an effective IC-POVM $\tilde{\boldsymbol \mu}=\Phi^\dagger(\boldsymbol \nu)$ on the input qubits. The reconstruction performances of QELM are affected by the choice of $\boldsymbol{\nu}$ and by the scrambling properties of the interaction $\mathcal{U}$ between the input qubits and the ancillae \cite{xiong2023fundamental}. It is common in the literature to assume the measurement $\boldsymbol{\nu}$ to act only on the reservoir. However, it is evident from our discussion that this is not necessary. In the following, we will assume the full set of input qubits and ancillae to be measured in the computation basis. In this way, the number of ancillae required to generate an effective IC-POVM is minimized.

\subsection{Expressivity and QELM}
\label{subsec:kernel}
Different choices of embedding of the classical input may influence the capacity of the QELM to reconstruct the functional relation between the elements of the training set. In order to understand how this choice may affect the performances, it is useful to observe that the QELM paradigm shows some similarities with the quantum kernel method, a feature that seems to be common to all quantum supervised learning routines \cite{xiong2023fundamental,Schuld2021}. In other words, we embed the classical data $\boldsymbol x$ into an element of a higher dimensional space, $\rho_{\boldsymbol x}\in\mathcal{D}$. The input Hilbert space is endowed with an inner product defining the kernel of the model:
\begin{equation}
    \kappa(\boldsymbol x,\boldsymbol y)=\Tr\left(\rho_{\boldsymbol x} \rho_{\boldsymbol y}\right)\,.
\end{equation}
Different kernels $\kappa(\boldsymbol x, \boldsymbol y)$ allow to approximate different kinds of functions. For instance, a popular choice is the \emph{rotation encoding}:
\begin{equation}
    U(\boldsymbol x)=\prod_{i=1}^{d_{\rm in}}\exp\left(-\frac{i}{2}x_i\sigma^{(i)}_x\right)\,,
\end{equation}
where $\sigma^{(i)}_x$ is the Pauli matrix $x$ acting on the $i$-th input qubit. The corresponding kernel is the \emph{translation invariant squared cosine kernel} \cite{Schuld2021}:
\begin{equation}
    \kappa(\boldsymbol x,\boldsymbol y)=\prod_{i=1}^{d_{\rm in}}|\cos(x_i-y_i)|^2
\end{equation}
The rotation embedding can be considered as a particular instance of the more general \emph{Fourier embedding}:
\begin{equation}
\label{eq:Fourier_Embedding}
    U(\boldsymbol x)=e^{-x_{\mathcal{X}}G}W^{(\mathcal{X})}\dots W^{(2)}e^
{-x_1G}W^{(1)}
\end{equation}
where $W^{(1)},W^{(2)},\dots$ are random unitary matrices acting on the input qubits, and $G$ is a Hermitian operator, for instance, a Pauli string. The name of this embedding comes from the peculiar form of the associated kernel:
\begin{equation}
\label{eq:fourier_kernel}
    \kappa(\boldsymbol x, \boldsymbol y)=\sum_{\boldsymbol n,\boldsymbol m}\alpha_{\boldsymbol n,\boldsymbol m}e^{i(\boldsymbol n\cdot \boldsymbol x-\boldsymbol m\cdot \boldsymbol y)}
\end{equation}
where $c_{\boldsymbol m,\boldsymbol n}$ are complex coefficients and $\boldsymbol n,\boldsymbol m\in \mathbb{\mathbb{R}}^{\mathcal{X}}$. The Fourier frequencies $\boldsymbol{m}$ are proportional to the gaps $\lambda_i-\lambda_j$, where $\lambda_{i,j}$ are eigenvalues of $G$ \cite{schuld2021effect}. The number of distinct Fourier frequencies is bounded from above by $2^{2N-1}-1$, where $N$ is the number of encoding qubits. As the number of frequencies increases, the model becomes more and more expressive. For this reason, studying the kernel associated with the embedding is a useful way to assess expressiveness \cite{xiong2023fundamental,Schuld2021,jager2023,jerbi2023}.  As shown in \cite{schuld2021effect}, the Fourier encoding models are universal approximators for appropriate choices of generator $G$.

\begin{figure}[tb]
    \centering
    \includegraphics[scale=0.2]{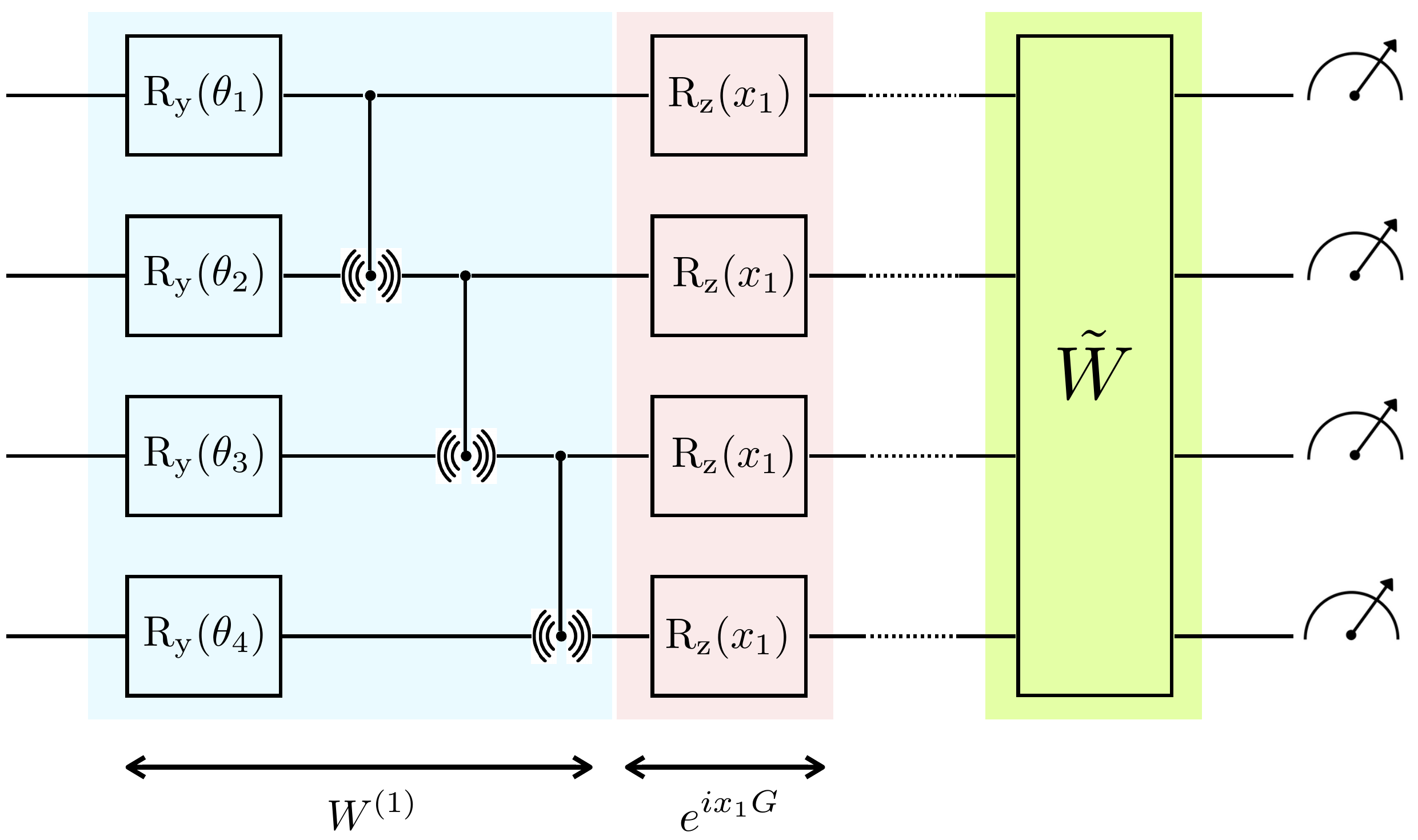}
    \caption{The QELM setup used for PES prediction with $4$ input qubits. The mixing unitary $\tilde W$ is made of three blocks, each of the same form as $W^{(1)}$, with random rotation angles drawn from the same ensemble. The two-qubit entangling gates in each layer are ECR gates.
    }
    \label{fig:basic_entangle}
\end{figure}

\begin{figure}[tb]
    \centering
    \includegraphics[scale=0.18]{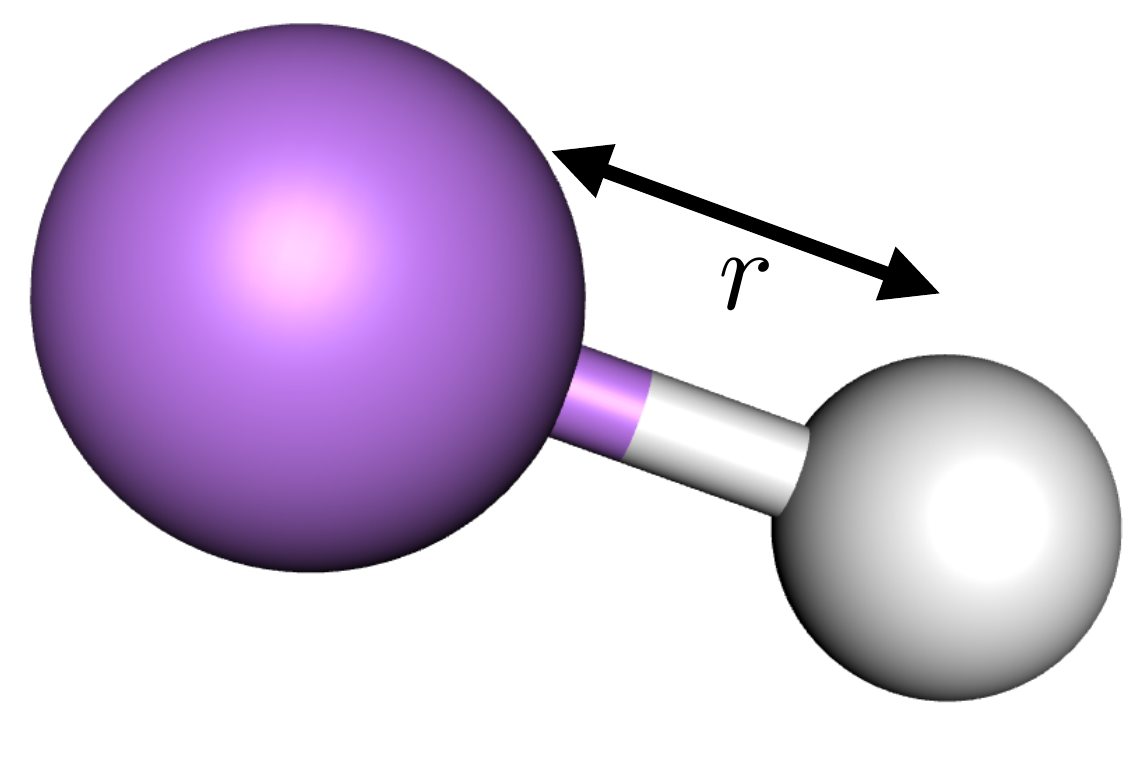}
    \caption{Parameterization of the geometry of a single molecule of lithium hydride.
    }
    \label{fig:LiH_geo}
    \includegraphics[scale=0.18]{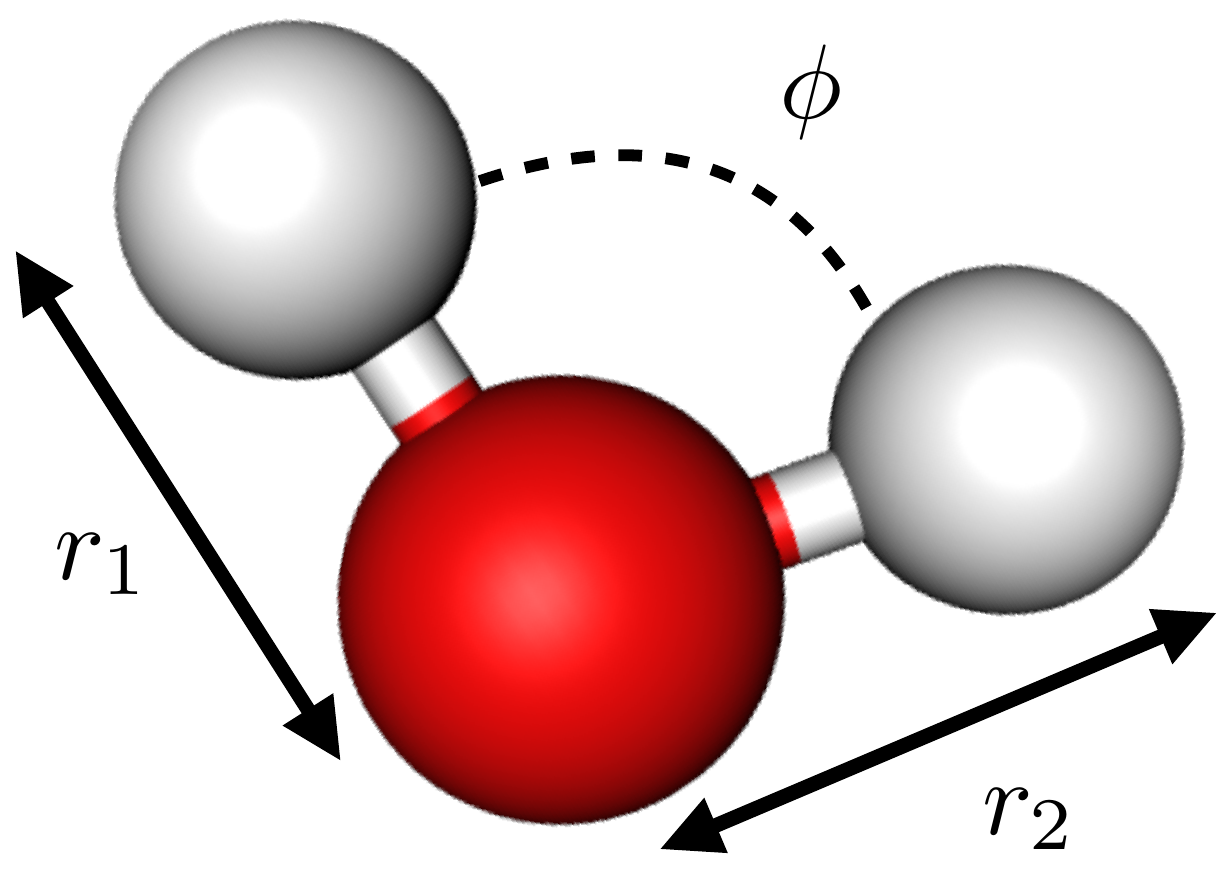}
    \caption{Parameterization of the geometry of a single molecule of water.
    }
    \label{fig:h2o_geo}
    \centering
    \includegraphics[scale=0.18]{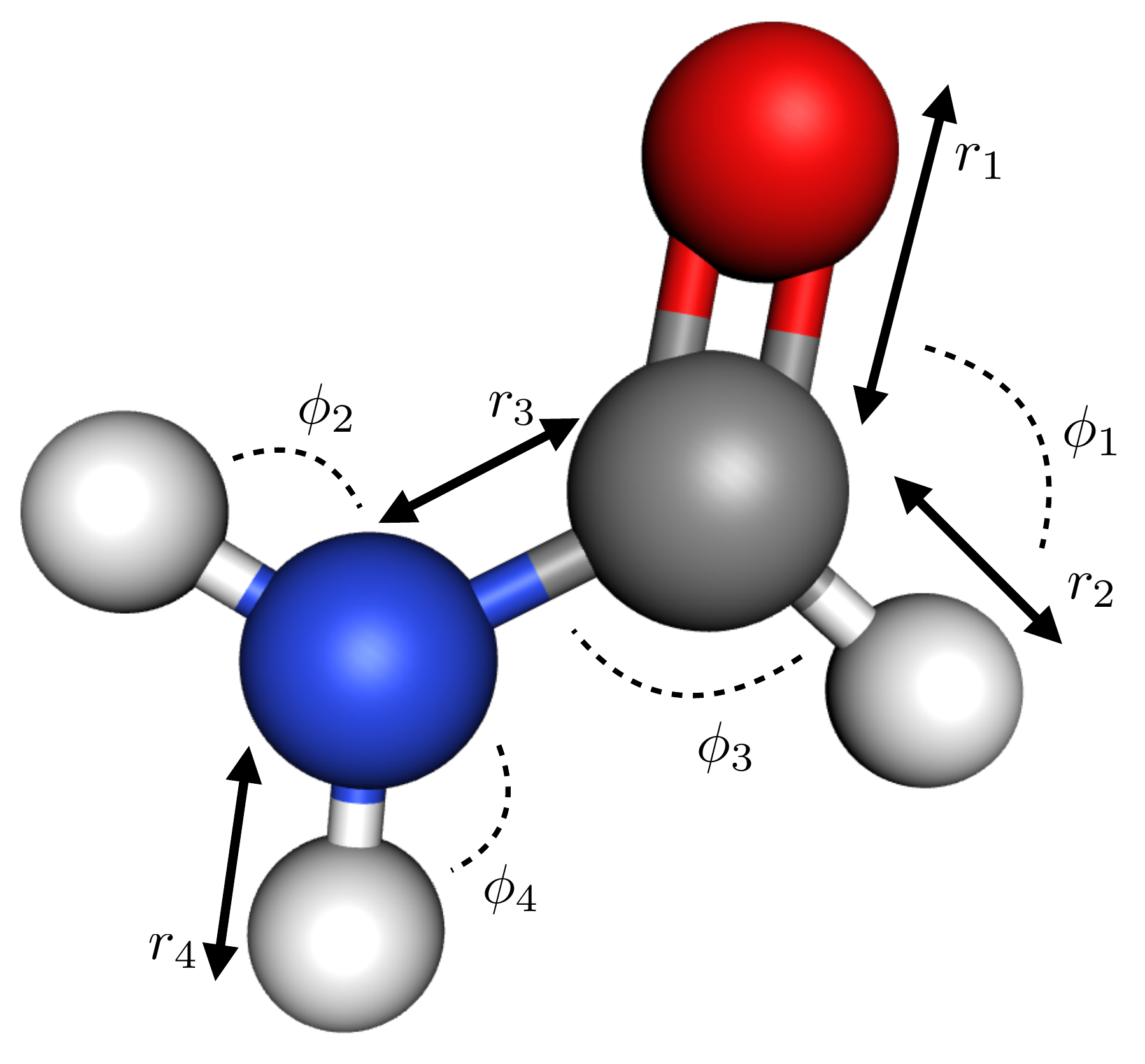}
    \caption{Parameterization of the geometry of a single molecule of formamide. The two ${\rm NO}$ bond lengths are taken equal and parameterized by the same coordinate $r_4.$
    }
    \label{fig:hconh2_geo}
\end{figure}

\section{PES predictions with QELM: simulations}
\label{sec:PesSimulations}
In this section we train a QELM to learn the PES and the FF for three different molecules of increasing complexity: lithium hydride (${\rm LiH}$), water (${\rm H}_2{\rm O}$) and formamide (${\rm HCONH}_2$). The datasets used during the training are generated in the same way for all the investigated molecules and their generation is discussed in greater detail in \sref{subsec:datasets}. The QELM model used can be adapted to molecules with an arbitrary number of atoms and is presented in \sref{subsec:QELM_setup}.

\subsection{Datasets}
\label{subsec:datasets}
The conformations of any molecule are represented in terms of generalized coordinates, either bond lengths or bond angles, that are usually used in the Z-matrix description of any species. The lithium hydride conformations are described in terms of a single coordinate, the bond length (\Fref{fig:LiH_geo}). The conformations of a water molecule are parameterized by three coordinates: the ${\rm HO}$ bond lengths $r_1,r_2$ and the oxygen and the ${\rm HOH}$ bond angle $\phi$ (\Fref{fig:h2o_geo}). The parameterization of formamide deserves a few more words. Such organic molecule is a particular instance of amides, the smallest molecules presenting a peptide bond. They have particular biological relevance: proteins are polymers whose units are connected by amide groups. The distinctive strength of the peptide (double) bonds forces the formamide to be a planar molecule in its ground state configuration. For such reason, we parameterized a formamide molecule assuming the geometry to be planar and the two ${\rm NH}$ bond lengths are assumed to be the same. The remaining conformations are parameterized in terms of $8$ generalized coordinates (instead of $12$) and their choice is shown in \Fref{fig:hconh2_geo}. 

The training and test configurations are then obtained by sampling random tuples of generalized coordinates in given intervals. In the three cases under consideration, these intervals are centered around the ground state configuration in order to focus on the geometries that are typically explored during a molecular dynamics simulations. In fact, for our three molecular species, the PES are characterized by a deep potential well which it's hard to escape. Given that the PES of ${\rm LiH}$ is actually a one-dimensional curve, in this case, we explore a wider part of the configuration landscape sampling the bond lengths from a larger interval than ${\rm H}_2{\rm O}$ or ${\rm HCONH}_2$. The electronic structure calculations performed to generate the used dataset were conducted using the PySCF software implemented in Pennylane package (${\rm LiH}$) or Gaussian 16 software \cite{g16} (${\rm H}_2{\rm O}$ and ${\rm HCONH}_2$). Since the main focus of this work is to assess the agreement of QELM with generic PES data obtained from traditional quantum chemistry methods, we are not restricted to a specific theoretical framework. For this reason, the level of theory chosen to investigate the different chemical species changes as a consequence of both the complexity of the molecule and the necessity of a properly sized dataset.

In the case of lithium hydride and water,  calculations were performed using DFT with M06 exchange-correlation functional \cite{zhao2008m06} and STO-3G (cc-pVDZ) basis set. For ${\rm LiH}$, the generated dataset is generated performing $170$ single point calculation on as many geometries, obtained by choosing the bond length values randomly in the range 0.9\AA~$\leq r\leq$ 4.5\AA. The QELM is trained over $50$ geometries while $120$ configurations are left for test. For ${\rm H}_2{\rm O}$, the dataset is generated performing $900$ single point calculations on as many geometries. Each geometry is obtained by choosing each value for the three internal coordinates at random, in a range centered on their respective equilibrium values, i.e. 0.964 $\pm$ 0.2\AA~for distances, and 102.792$^{\circ}$ $\pm$ 13.0$^{\circ}$ for the ${\rm HOH}$ angle. The QELM is trained over $300$ elements of the whole dataset, while $600$ configurations are left for test.
In the formamide case, the dataset consists of approximately 6500 structures (even if much less are actually used for the training). Each of the geometries that constitute the dataset was generated by randomly sampling values for the aforementioned bonds and angles within specific intervals. In particular, these intervals are centered on the value of the respective internal equilibrium coordinates, with ranges of: i) $\pm0.10$ \AA~ for bonds involving a hydrogen atom; ii) $\pm0.15$ \AA~ for the remaining bonds; and iii) $\pm8.00^\circ$ for all angles. For each structure, a calculation was performed at the HF/SVP level of theory to obtain the electronic energy and atomic force values.

\begin{table}[t]
    \small
    \begin{tabular}{@{}c|| c| c|c|c|c}
    \br
        \cellcolor{LightRed}${\rm LiH}$& \cellcolor{LightRed} $\,\,$ \# shots$\,\,\,$ & \cellcolor{LightRed}$\quad$ RMSE(E) $\quad$& \cellcolor{LightRed}$\quad$ RMSE($F_1$) $\quad$& \cellcolor{LightRed} $\,\,$\# qubits$ \,\,$ & \cellcolor{LightRed}$\,\,$ depth $\,\,$ \\
        \mr
        statevector& $\infty$ & $6.9\cdot 10^{-5}$ & $ 4.2\cdot 10^{-4} $  &  $5$ & 27\\
        & & & & &
        \\
        QASM simulator& $4\cdot 10^{4}$ & $9.3\cdot 10^{-3} $ & $1.8\cdot 10^{-2} $ & $4$ & 26 \\
        & & & & &
        \\
        IBM\_BRISBANE& $4\cdot 10^{4}$ & $3.2\cdot 10^{-2}$ & $5.0\cdot 10^{-2}$ & $4$ & 26 \\
        \br
        \cellcolor{LightRed}${\rm H}_2{\rm O}$& \cellcolor{LightRed} $\,\,$ \# shots$\,\,\,$ & \cellcolor{LightRed}$\quad$ RMSE(E) $\quad$& \cellcolor{LightRed}$\quad$ RMSE($F_1$) $\quad$& \cellcolor{LightRed} $\,\,$\# qubits $\,\,$ & \cellcolor{LightRed}$\,\,$ depth $\,\,$ \\
        \mr
        statevector & $\infty$ & $2.0\cdot 10^{-6}$ & $ 9.3\cdot 10^{-6}$  &  $7$ & 41\\
        & & & & &
        \\
        QASM simulator & $2\cdot 10^{4}$ & $3.6\cdot 10^{-3}$ & $1.2\cdot 10^{-2}$ & $5$ & 39 \\
        & & & & &\\
        IBM\_BRISBANE & $2\cdot 10^{4}$ & $5.2\cdot 10^{-3}$ & $2.2\cdot 10^{-2}$ & $5$ & 39 \\
        \br
         \cellcolor{LightRed}${\rm HCONH}_2$& \cellcolor{LightRed} $\,\,$ \# shots$\,\,\,$ & \cellcolor{LightRed}$\quad$ RMSE(E) $\quad$& \cellcolor{LightRed}$\quad$ RMSE($F_1$) $\quad$& \cellcolor{LightRed} $\,\,$\# qubits $\,\,$ & \cellcolor{LightRed}$\,\,$ depth $\,\,$ \\
        \mr
        statevector& $\infty$ & $3.4\cdot 10^{-4}$ & $ 9.7\cdot 10^{-4}$  & $9$ & 73\\
        & & & & &
        \\
        quasm simulator& $2\cdot 10^{4}$ & $7.0\cdot 10^{-3}$ & $2.1\cdot 10^{-2}$ & $7$ & 71\\
        & & & & &
        \\
        IBM\_BRISBANE& $2\cdot 10^{4}$ & $1.3\cdot 10^{-2}$ & $4.4\cdot 10^{-2}$ & $7$ & 71\\
        \br
    \end{tabular}
    \caption{\label{tab:results} Summary of the performances of the QELM trained to learn PESs and FFs of a molecule of ${\rm LiH}$, ${\rm H}_2{\rm O}$ and ${\rm HCONH}_2$. The RMSE is measured in ${\rm Ha}$ for the energy and ${\rm Ha}/$\AA~  for the force.}
\end{table}

\subsection{QELM setup}
\label{subsec:QELM_setup}
The gate parameters in quantum hardware are usually rotation angles. For this reason, we need to perform a rescaling of the coordinate, with particular attention to radial coordinates. Each bond length is rescaled with respect to a reference one $\overline{r}$. The reference distances are chosen to be $\overline{r}_{{\rm LiH}}=6$\AA$\,$ and $\overline{r}_{{\rm H}_2{\rm O}}=\overline{r}_{{\rm HCONH}_2}=2$\AA. The bond angles are instead rescaled by a factor $2$. The input data of water are thus of the following form:
\begin{equation}
    \boldsymbol{x}=\left\{\frac{r_1}{\overline{r}}\pi\,,\,\frac{r_2}{\overline{r}}\pi\,,\,\frac{\phi}{2}\right\}\,.
\end{equation} 
with analogous expressions for lithium hydride and formamide.
The string $\boldsymbol x$ is fed in the state of $N$ qubits using a Fourier encoding \eref{eq:Fourier_Embedding} with:
\begin{equation}
\label{eq:G_H2O}
G=\frac{1}{2}\sum_{i=1}^{N}\sigma_{z}^{(i)}
\end{equation}
This amounts to act on each input qubit with an $R_z$ gate. Observe that the number of input qubits of the encoding is not fixed and can be increased at wish in order to improve performances. The random unitaries $W^{(1)},W^{(2)},\dots$ in the Fourier encoding are made of one layer of the form depicted in \Fref{fig:basic_entangle}, where $\theta_i$ are random rotation angles uniformly drawn in the interval $(0,\pi/2)$. All the input qubits are measured in the computational basis, with no addition of ancillae. Let us also stress that the Fourier encoding as presented in \eref{eq:Fourier_Embedding} can be adapted to any molecule or system of molecules with an arbitrary number of generalized coordinates.

The kernel associated with the choice of generator \eref{eq:G_H2O} is a Fourier kernel \eref{eq:fourier_kernel} with integer frequencies $-N,-N+1,\dots,N-1,N$. Even if the number of frequencies scales only linearly with the number of qubits, it must be stressed that the choice of embedding and gates in the setup is driven by the wish to minimize the depth of the quantum circuit once it is implemented on real quantum hardware. In fact, as it will be clear in the following, the accuracy of the PES predictions grows as the number of encoding qubits increases, reflecting the wider spectrum of Fourier frequencies of the model (see \Fref{fig:scaling}). 

In the following, we will present the actual experimental implementation of the proposed setup on the IBM\_{BRISBANE} device, a quantum processor with $127$ superconducting transmon qubits. Such a device admits thee native single-qubit gates ($X$, $\sqrt X$ and ${\rm R_z}(\phi)$ for any angle) and a single two-qubit gate, the \emph{Echoed cross-resonance} gate (ECR):
\begin{equation}
    {\rm ECR}_{12}={\rm R_{xx}}(\pi/2){\rm X}_2
\end{equation}
The ECR gate is equivalent to the CNOT gate up to single-qubit rotations. Any other gates must be decomposed as a combination of native gates and this may cause a huge increase in the circuit depth. For instance:
\begin{eqnarray}
    {\rm R_x}(\phi)&={\rm R_z}\left(\frac{\pi}{2}\right)\sqrt{X}\,{\rm R_z}(\pi+\phi)\,\sqrt{X}\,{\rm R_z}\left(\frac{5\pi}{2}\right)\\
    {\rm R_y}(\phi)&={\rm R_z}(-\pi)\,\sqrt{X}\,{\rm R_z}(\pi-\phi)\,\sqrt X
\end{eqnarray}
Since the ${\rm R_y}$ gates are slightly cheaper in terms of native gates, they should be preferred in order to minimize the circuit depth. For this reason, our circuit is built out of ${\rm R_{z,y}}$ and ECR gates only.

\subsection{Results}
\label{subsec:res}
For each atomic species that we considered, we performed three different kinds of analysis: first, we simulated the performances of the QELM in the ideal case of infinite statistics and absence of noise. We refer to this case as \emph{statevector} simulation. Then we reduce the level of approximation simulating the QELM training with finite statistics but in noiseless conditions. To address this task, we use the \emph{QASM simulator} of the Qiskit package \cite{Qiskit}. Finally, we implement the QELM setup on IBM\_BRISBANE, taking thus into account also the actual noise present on NISQ devices. It must be stressed that we do not perform any error mitigation on the quantum-hardware output data.

\begin{figure*}[h!]
    \includegraphics[width=\linewidth]{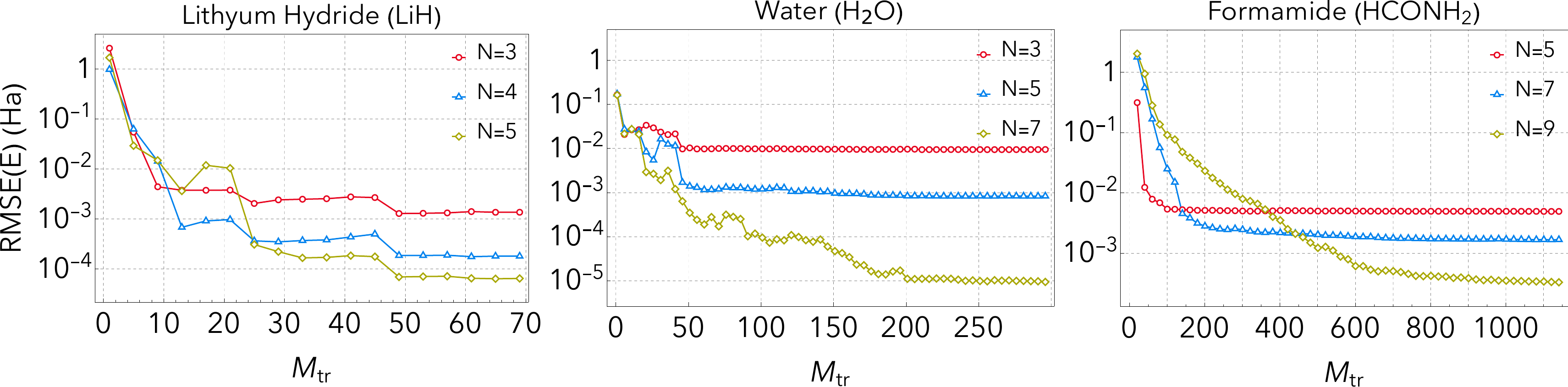}
    \caption{ Behaviour of the energy RMSE in statevector simulations for a different number of the encoding qubits and different sizes of the training set.}
    \label{fig:scaling}
\end{figure*}

The statevector simulation is the ideal playground to test the achievable precision of the QELM prediction when the parameters of the setup are varied. The only freedom left in our configuration is in the number of encoding qubits and the dimension of the training set. In \Fref{fig:scaling} we plot the behavior of the statevector RMSE for the energy for all the molecules under investigation when both $N$ and $M_{\rm train}$ are varied. As one can observe, as the number of qubits increases, the performance of the QELM improves. This is consistent with the idea that a larger reservoir can be used to approximate a wider range of functions \cite{xiong2023fundamental}. What is interesting is that the RMSE always reaches a plateau above a certain dimension of the training set. Above this threshold value, any further information does not significantly affect the accuracy of the predictions on the test set. The reason relies on the linear regression that is carried out in post-processing in the QELM paradigm. As explained in \Fref{sec:QEML}, the training consists of a single linear regression once the matrix
\begin{equation}
    W=\boldsymbol Y_{\rm tr}\boldsymbol\,P_{\rm tr}^+
\end{equation}
is determined. The reservoir dependence of the training is all contained in the probability matrix $\boldsymbol P_{\rm tr}^+$. It is well known that some information about the accuracy of the training can be extracted by looking at the singular values of the probability matrix only \cite{innocenti2023potential, higham2002accuracy}. However, the number of singular values of  $\boldsymbol P_{\rm tr}^+$ is always less than $\min(2^{N},M_{\rm tr})$. This implies that, as long as the number of qubits is kept fixed, the larger the value of $M_{\rm tr}$ is compared to $2^{N}$, the less any new training element will impact performance. Looking at \Fref{fig:scaling}, the RMSE plateau is reached for:
\begin{equation}
2^{N}<M_{\rm tr}\lesssim 2^{N+1}\,.
\end{equation}
In the first row of each sub-table of \Tref{tab:results}, we collect the results of training for the three atomic species for a specific choice of $N$. The RMSE of the energy is in all three cases well below the typical energy fluctuations in the dataset, estimated by the square root of the variance, $\sqrt{{\rm var}(\boldsymbol E_{\rm tr})}$, where $\boldsymbol E_{\rm tr}$ is the vector whose elements are the training energies. In all the three cases, ${\rm LiH}$, ${\rm H}_2{\rm O}$ and ${\rm HCONH}_2$, the variance is of order $2\cdot{10}^{-2}\,{\rm Ha}$. For the force associated to bond lengths instead, $\sqrt{{\rm var}(\boldsymbol F_r)}\approx 10^{-1}{\rm Ha}/$\AA. 

It is interesting to compare these results with the performances (in the infinite-statistics, noiseless ideal case) of the VQE routine proposed in \cite{kiss2022quantum}, where an analogous study of ${\rm LiH}$ and ${\rm H}_2{\rm O}$ is carried out \Tref{tab:vqe}.
\begin{table}[h!]
\centering
\caption{\label{tab:vqe}Performances (statevector simulation) of the VQE routine proposed in \cite{kiss2022quantum} to learn PES and FF of LiH and H$_2$O, together with the classical machine learning routines used as benchmark.}
\begin{tabular}{c||c|c|c|c}
\br
        \cellcolor{LightRed}${\rm LiH}$& \cellcolor{LightRed} $\quad$ RMSE(E) $\quad$& \cellcolor{LightRed} $\,\,$\# qubits$\,\,$ & \cellcolor{LightRed}$\,\,$ depth $\,\,$ & \cellcolor{LightRed} \# of parameters \\
        \mr
         VQE & $1.4\cdot 10^{-4}\,{\rm Ha}$ & $3$ & 682 & 73 \\
            DNN & $1.5\cdot 10^{-3}\,{\rm Ha}$ & $\times$ & $\times$ & 73
         \\
        \br
        \cellcolor{LightRed}${\rm H}_2{\rm O}$& \cellcolor{LightRed} $\quad$ RMSE(E) $\quad$ & \cellcolor{LightRed} $\,\,$\# qubits$\,\,$ & \cellcolor{LightRed}$\,\,$ depth $\,\,$& \cellcolor{LightRed} \# of parameters\\
        \mr
        VQE & $1.8\cdot 10^{-4}\,{\rm Ha}$ & $3$ & 818 & 87\\
        DNN & $2.2\cdot 10^{-4}\,{\rm Ha}$ & $\times$ & $\times$ & 87\\
        BPNN & $2.5\cdot 10^{-5}\,{\rm Ha}$ & $\times$ & $\times$ & 1642
        \\
        \br
\end{tabular}
\end{table}

In both cases, the QELM outperforms the VQE, not only in terms of RMSE but also in terms of needed resources. In fact, even if the VQE routine of \cite{kiss2022quantum} uses a smaller circuit ($3$ qubits), it needs a circuit with a depth approximately $200$ times bigger than in the QELM counterpart. Moreover, the VQE circuit needs to be run many more times, in order to obtain the gradients of the parameters (see \sref{app:shift_rule}) and in order to perform more training epochs.

In \Tref{tab:vqe}, we also included for completeness the comparison of the study conducted by the authors of \cite{kiss2022quantum} using a fully connected multi-layered deep neural network (DNN)  with hyperbolic tangent activation function (five layers with $[7, 4, 5, 2, 1]$ units for LiH and 6 layers with $[7, 4, 6, 2, 2, 1]$ units for H$_2$O) and the Behler-Parrinello-like neural network (BPNN) of \cite{singraber2019library}. The number of parameters to be optimized is taken to be the same for VQE and DNN for each molecular species while it is greater for BPNN. As it is evident, also the VQE can outperform the classical networks. The QELM not only improves the VQE performances, but it does so while keeping minimal quantum computational cost.

\begin{figure*}[h!]
    \includegraphics[width=\linewidth]{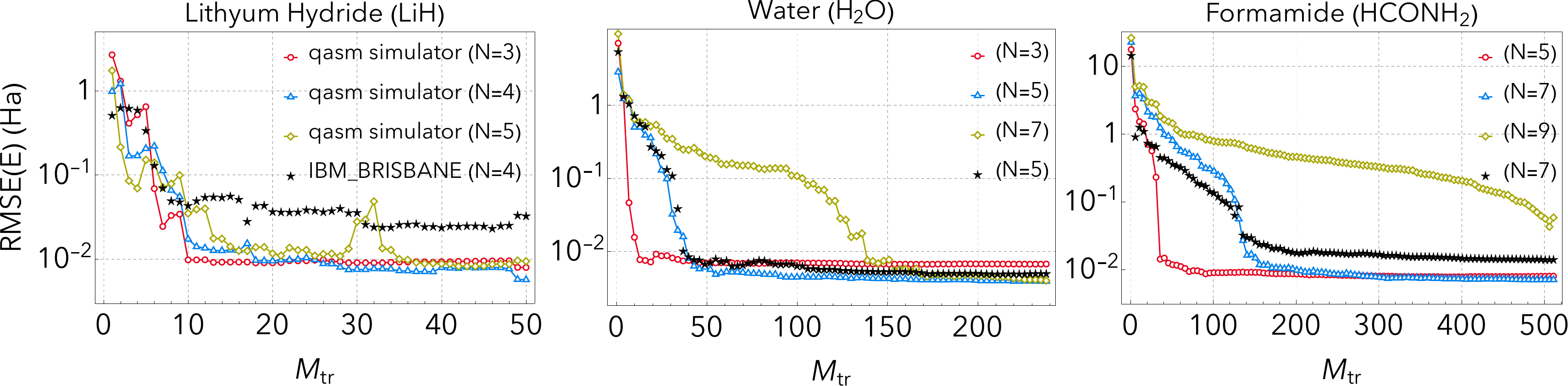}
    \caption{ Behaviour of the energy RMSE in QASM simulations and IBM\_BRISBANE realizations for different numbers of the encoding qubits and different sizes of training set.}
    \label{fig:stat_scaling}
\end{figure*}

In realistic implementations, the probability matrices $\boldsymbol P_{\rm tr}$ and $\boldsymbol P_{\rm test}$ are estimated with finite precision, due to the fact that probabilities are estimated by performing a limited number of shots on a quantum device. This implies that the accuracy of the QELM predictions is in general lower than the ideal case. The results of the training either in the noiseless simulation or the real-hardware implementation on IBM\_BRISBANE are again collected in table \Tref{tab:results}, where they can be compared with their statevector counterpart. One can see that because of the finite statistics, the RMSE increases by a factor that depends on the specific molecule. However, the RMSE in the QASM simulations and IBM\_BRISBANE implementation are always of the same order of magnitude, meaning that the noise does not affect significantly the performances and the noiseless simulations are reliable. The dependence of the RMSE on the number of training elements follows the same argument as above. Moreover, looking at \Fref{fig:stat_scaling} it is evident that another effect of the finite statistics is that the RMSEs for different numbers of qubits are much closer to each other on the plateau. 

\begin{figure*}[h!]
    \includegraphics[width=\linewidth]{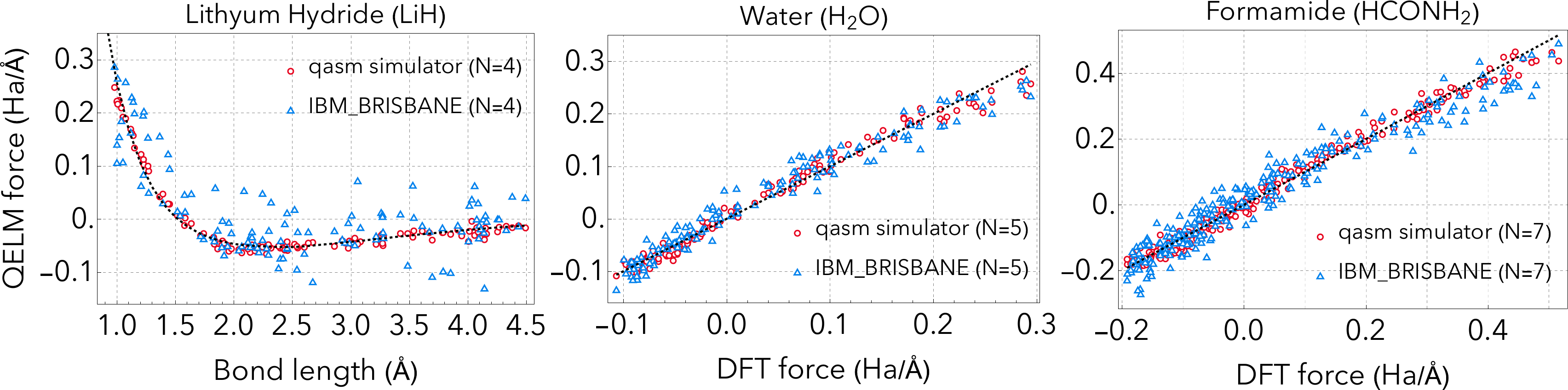}
    \caption{ Comparison between classical forces and QELM predictions obtained with QASM simulations and IBM\_BRISBANE realizations. In the ${\rm LiH}$ case, we explicitly plot the force field curve.}
    \label{fig:curves}
\end{figure*}

In order to have a visual perspective of the accuracy of our predictions at finite statistics (either in QASM simulation or IBM\_BRISBANE realization), we plot in \Fref{fig:curves}, the value of the force computed with classical methods against the respective QELM prediction (in the ${\rm LiH}$ case we rather plot the value of the forces in the $r-F$ plane).


\section{Conclusions}
In this work, we have demonstrated a fits application of QELM to 
quantum chemistry. The QELM is trained to learn the functional dependence of the potential energy surface and force fields of a molecule on a set of generalized coordinates that parameterize its possible conformations. The predictions of the QELM can be then exploited to speed up ab-initio molecular dynamics computations that require the knowledge of the PES and FF at HS or DFT precision.

We have studied in detail three case studies, ${\rm LiH}$, ${\rm H}_2{\rm O}$ and ${\rm HCONH}_2$ but the proposed setup is scalable, in the sense that can be applied to molecules with any number of atoms and degrees of freedom. The setup is also optimized for practical implementation on quantum devices, with IBM superconducting transmon quantum computers in mind. However, by changing the set of native gates, the setup can be easily adapted to other platforms, e.g. trapped-ion. The precision of the prediction can be arbitrarily increased by utilizing more encoding qubits. The number of qubits in turn fixes the dimension of the training set that should be used.

The accuracy of the QELM predictions is first tested in noiseless simulations at infinite statistics, finding an exceptional agreement. We then investigate the performances at finite statistics, either with noiseless simulation and also with practical implementations on the IBM\_BRISBANE quantum device. In both cases we find encouraging results, suggesting that the proposed method can be effectively competitive in a near future. It must be stressed that QELM is particularly efficient in terms of quantum resource needed (it does not require any optimization of quantum circuit parameters via gradient descent) and in terms of classical post-processing (we do not perform any error mitigation).

We leave to future work the application to more complicated molecules in order to understand the ultimate frontiers of our proposal and the study of efficient methods to overcome the limitations deriving from finite statistics.

\ack

We would like to thank Luca Innocenti for invaluable discussions. We acknowledge the use of IBM Quantum Credits for this work. The views expressed are those of the authors and do not reflect the official policy or position of IBM or the IBM Quantum team. GLM and GMP acknowledge funding from the European Union - NextGenerationEU through the Italian Ministry of University and Research under PNRR - M4C2-I1.3 Project PE-00000019 "HEAL ITALIA" (CUP B73C22001250006 ). 
MB thanks ``SiciliAn MicronanOTecH Research And innovation CEnter - SAMOTHRACE'' (MUR, PNRR-M4C2, ECS0000002), spoke 3, Università degli Studi di Palermo, for financial support.
SL and GMP  acknowledge support by MUR under PRIN Project No. 2022FEXLYB Quantum Reservoir Computing (QuReCo) and by the “National Centre for HPC, Big Data and Quantum Computing (HPC)” Project CN00000013 HyQELM – SPOKE 10.

\section*{References}
\bibliographystyle{iopart-num}
\bibliography{bibliography}

\appendix

\section{Force field generation with quantum routines}
\label{app:shift_rule}
All supervised QML routines consist in evaluating the expectation values of some operator $\Tr(\mathcal{O}(\boldsymbol\theta,\boldsymbol x)\rho_0)$ where $\rho_0$ is some reference state. $\boldsymbol \theta$ is a collection of parameters that in VQE routines gets optimized to fit the training data. The optimization consists of a gradient descend algorithm performed on a classical computer, while the gradients of the parameters are estimated on a quantum computer through the parameter-shift rule \cite{mitarai2018quantum,schuld2019evaluating}:
\begin{equation}
\label{eq:shift_rule}
    \partial_\theta f(\theta)\,=\,\,\lambda\left(f\left(\theta+\frac{\pi}{4\lambda}\right)-f\left(\theta-\frac{\pi}{4\lambda}\right)\right)\,,
\end{equation}
where $f(\theta)=\Tr(e^{-i\theta G}\mathcal{O}e^{i\theta G}\rho)$ and $G$ is a Hermitian operator whose spectrum has only two distinct eigenvalues $\pm\lambda$. When $G$ has more than two distinct eigenvalues, it is possible to construct generalized shift rules as in \eref{eq:shift_rule} where the sum runs over more than two shifts. Using the parameter-shift rule, it is also possible to compute the derivative with respect to the input data. In the context of PES predictions, where the VQE is used to predict conformational energies, these derivatives coincide with the forces. This knowledge can be used to train a unique circuit able to provide either energies and forces. However, in order to get all this information, the circuit needs to be run $2\mathcal{\chi}\mathcal{S}+1$ times, where $\mathcal{\chi}$ is the number of coordinates and $\mathcal{S}$ is the number of shifts appearing in the parameter-shift rule. 

The parameter-shift rule can be also applied in the QELM setup but this does not lead to great advantages. In fact, in the QELM routines, a unique encoding of the coordinates $\boldsymbol x$ is enough to predict either energies and forces: the unique difference is that a bigger matrix $\boldsymbol W$ is obtained at the classical stage of the algorithm. This greatly reduces the running time of a QELM simulation for force field predictions.

\end{document}